\documentclass[twocolumn,showpacs,showkeys,preprintnumbers,amsmath,amssymb]{revtex4}
\usepackage{graphicx}
\usepackage{epsfig}
\usepackage{color}

\newcommand{\beq}{\begin{equation}}
\newcommand{\eeq}{\end{equation}}
\newcommand{\bea}{\begin{eqnarray}}
\newcommand{\eea}{\end{eqnarray}}
\newcommand{\bwd}{\begin{widetext}}
\newcommand{\ewd}{\end{widetext}}

\def    \Sm{{\mathbb{S}}}

\begin{document}

\preprint{BNL-71394-2003-JA} \preprint{SLAC-PUB-10004}

\title{Coupled-Bunch Beam Breakup due to Resistive-Wall Wake\footnote{Work 
supported by Department of Energy contract DE--AC02--98CH10886 (Jiunn-Ming 
Wang) and contract DE--AC03--76SF00515 (Juhao Wu)}}

\author{Jiunn-Ming Wang}
\email{jmwang@bnl.gov}
\affiliation{National Synchrotron Light Source, Brookhaven National
Laboratory, Upton, NY 11973}
\author{Juhao Wu}
\email{jhwu@SLAC.Stanford.EDU}
\affiliation{Stanford Linear Accelerator Center, Stanford University, Stanford,
CA 94309}

\date{\today
\\ ,,Submitted to Physical Review E}

\begin{abstract}
The coupled-bunch beam breakup problem excited by the resistive
wall wake is formulated. An approximate analytic method of finding
the asymptotic behavior of the transverse bunch displacement is
developed and solved.
\end{abstract}

\pacs{29.27.Bd; 52.35.Qz; 41.75.Ht; 07.85.Qe} \keywords{Coherent
effect; Collective effect; Beam Breakup} \maketitle

\section{Introduction}
The coupled-bunch beam breakup (bbu) problem in a periodic linac
excited by the resonance wake is well understood
\cite{GCC85,DW88,B&K}. However, there are no systematic studies
for the corresponding problem excited by the resistive wall
impedance. This study of the resistive wall bbu problem is
necessitated by the recently proposed PERL project \cite{Perl01}.
For PERL, the light source consists of twelve undulators, each
twelve meters long, totally 144 meters. The beam is shielded from
the environment by circular copper pipes of a very small radius
$b$=2.5 mm. The proposed injection cycle is twelve hours. It is
crucial to know if the PERL beam can survive the bbu. We present
our theoretical results for the resistive-wall coupled-bunch beam
breakup problem in this paper. Some of the results obtained here
has been briefly reported in Ref. \cite{WW02}.

The paper is organized as follows: In Sec. \ref{SecEOM}, we set up
the equations of motion and then solve the related eigenvalue
problem. Physically, the eigenfunction so obtained describe the
beam coherent-oscillation of an ``extended problem''. In Sec.
\ref{SecInitial}, we give a formal solution for the initial value
problem. The solution consists of an integral representation for
the transverse position of the $M$-th bunch at a longitudinal
position $z$ in terms of the eigenfunctions obtained in the
previous section \cite{extended}. The asymptotic limit,
$M\rightarrow\infty$, of the transient solution is then obtained
in Sec. \ref{OneOffset} for two extreme cases: the No Focusing
(NF) case and the Strong Focusing (SF) case. In Sections
\ref{SecEOM} $\sim$ \ref{SecDamping}, we treat the case where only
one bunch is offset initially. While in Sections \ref{SecEOM}
$\sim$ \ref{OneOffset}, we treat the case where every bucket of
the linac is filled by the same amount of charge, we treat in
Section \ref{SecDamping} the case where the filling pattern is
such that the beam has periodically unfilled gaps. The results of
the section \ref{SecDamping} is compared to the results of the
preceding sections. The conclusion we draw from the comparison is
that the asymptotic resistive-wall coupled-bunch bbu is a locally
averaged current problem. In Section \ref{SecAll} we go back to
the problem where each bucket is symmetrically filled. The
difference between this section and the section \ref{OneOffset} is
that here we treat the case where initially the transverse
position of every bunch is offset by the same amount -- injection
error. By comparing the results of Section \ref{SecAll} with those
of Section \ref{OneOffset}, we observe ``Screen Effects'' for the
injection error case.

\section{Equation of Motion and the Eigenvalue Problem}\label{SecEOM}
An electron bunch train consisting of a series of identical point
like bunches passes through a circularly cylindrical pipe of
radius $b$ and conductivity $\sigma$. The entrance to the pipe is
located at $z = 0$, and the $M$-th bunch, $M = 0,1,2...$, moves in
the $z$ direction according to $z = c\,t - M\,c\,
\tau_{\mathrm{B}}$, where $\tau _{\mathrm{B}}$ is the bunch
separation in units of seconds. Inside the pipe, the equation of
motion for a particle in the bunch $M$ is
    \beq
    \hat{O}\,y_M\equiv y''_{\mathrm{M}} (z) + k^2_y y_{\mathrm{M}}
    (z) = \sum_{N=0}^{M-1}
    S(M-N)y_{\mathrm{N}}(z)\;  ,
    \label{eq:motion}
    \eeq
where the prime ``$\prime$'' stands for $d/dz$. The right hand
side of Eq. (\ref{eq:motion}) represents the effects of the wake
force , and for the resistive wall wake \cite{morton}
    \beq
    S(M)=a/\sqrt{M}  \;  ,
    \label{S}
    \eeq
with
    \beq
    a = 4\frac{I_{\mathrm{B}}}{I_{\mathrm{A}}}\frac{1}{b^3}
    \delta_{\mathrm{skin}}\;,
    \label{a}
    \eeq
where $I_{\mathrm{B}}=e\,N_{\mathrm{B}}/\tau_{\mathrm{B}}$,
$e\,N_{\mathrm{B}} =$ bunch charge, $I_{\mathrm{A}}\equiv 4\,
\pi\,\epsilon_o\,mc^3\gamma/e=\gamma\,I_{{\mathrm{Alfv}}
{\grave{\mathrm{e}}} {\mathrm{n}}}$, $I_{{\mathrm{Alfv}}
{\grave{\mathrm{e}}}{\mathrm{n}}} \approx 17,000$ Amp,
$\gamma$\,=\,the relativistic energy factor, and
$\delta_{\mathrm{skin}}=\sqrt{2/\mu_o\sigma\omega_{\mathrm{B}}}$\,=
the skin depth corresponding to the bunch frequency
$f_{\mathrm{B}}=\omega_{\mathrm{B}}/2\pi =1/\tau_{\mathrm{B}}$. We
ignore the effects of the wake force of a bunch on itself; as a
consequence, the upper limit of the sum in Eq. (\ref{eq:motion})
is $M-1$ instead of $M$. The thickness of the beam pipe is assumed
to be $\infty$ for convenience. Also notice that the bunch $N$ is
in front of the bunch $M$ if $M>N$.

In writing the above equations, we assumed the linac to be
uniformly filled. For such a case, the locally averaged current
$I_{\mathrm{average}}=I_{\mathrm{B}}$. For the case of non-uniform
filling, an example of that will be discussed in Section
\ref{SecDamping}, the equation (\ref{eq:motion}) has to be
modified.

The right hand side of Eq. (\ref{eq:motion}) is a convolution sum,
therefore, it can be diagonalized by a Fourier transform. Define
    \beq
    F(\theta)=\sum_{M=1}^{\infty}\frac{1}{\sqrt{M}}\,e^{iM\,\theta}\;  ,
    \label{Ftheta}
    \eeq
and
    \beq
    \xi (\theta ,z)= \sum_{M=0}^{\infty}y_{\mathrm{M}}(z)e^{iM\,\theta}\;  ,
    \label{eq:xi}
    \eeq
then
    \beq
    y_{\mathrm{M}}(z)=\frac{1}{2\pi}\int_{-\pi}^{\pi}d\theta\,e^{-iM\theta}\xi
    (\theta
    ,z)\; ,
    \label{xi_inv}
    \eeq
and
    \beq
    \xi ''(\theta ,z)+k^2_y\xi(\theta ,z)  = aF(\theta ) \xi
    (\theta ,z)\; .
    \label{xiequation}
    \eeq
The last equation is an eigenvalue equation, with the parameter
$\theta$ playing the role of distinguishing different eigenvalues.
For the coherent mode $\theta$, we see from Eq. (\ref{eq:xi}) that
the parameter $\theta$ is the phase difference of the adjacent
bunches in this mode. Recall that in a storage ring, a symmetric
coupling bunch mode $n$ is characterized by the Courant-Sessler
phase factor $\exp(i\,2\, \pi\,n/h)$ \cite{CS66}, where $h$ is the
number of the bunches in the ring. We can think of the phase
$\exp(i\theta)$ as the limit of the Courant-Sessler factor as both
$n$ and $h\rightarrow\infty$ while $n/h=\theta$ remains finite.
The eigenvalue for the mode $\theta$ is, from Eq.
(\ref{xiequation}),
    \beq
    k_c(\theta )=\sqrt{k^2_y - aF(\theta)} \; ,
    \label{kc}
    \eeq
and the corresponding eigenvetors are
    \beq
    \cos[k_c(\theta)z], \hspace{1 cm}\mbox{or} \hspace{1 cm}
    \sin[k_c(\theta)z]
    \label{eigenSol}
    \eeq

The function $F(\theta )$ can be written as \cite{zeta}
    \bea\label{F_expand}
    F(\theta)\;
    &=&\;
    \sqrt{\frac{i\pi}{\theta}}+\sum^{\infty}_{n=0}\zeta_{\mathrm{Riemann}}
    \left(\frac 12-n\right)\frac{(i\theta)^n}{n!}
    \nonumber\\
    &\approx&\;
    \sqrt{\frac{i\pi}{\theta}}-1.460-0.208\,i\,\theta+O(\theta^2)\;,
    \eea
where $\zeta_{\mathrm{Riemann}}(x)$ is the Riemann's Zeta
function. The function $F(\theta)$ has a branch point at $\theta
=0$, therefore, through Eq. (\ref{xiequation}), $\xi (\theta ,z)$
also has a singular point at the same position. Since Eq.
(\ref{xi_inv}) is the inverse of Eq. (\ref{eq:xi}) and we look for
$y_{\mathrm{M}}$ with $M>0$, causality requires this singularity
to lie below the contour of Eq. (\ref{xi_inv}) on the $\theta$
plane. In order to explain this point more clearly, let us
introduce
    \beq
    \zeta \equiv e^{i\theta}\; .
    \label{zeta}
    \eeq
In term of this variable, Eqs. (\ref{Ftheta}) $\sim$
(\ref{xiequation}) become
    \beq
    F(\zeta)=\sum_{M=1}^{\infty}\frac{1}{\sqrt{M}}\,\zeta^M\;  ,
    \label{Fzeta}
    \eeq
    \beq
    \xi (\zeta ,z)= \sum_{M=0}^{\infty}y_{\mathrm{M}}(z)\zeta^M \;,
    \label{eq:xizeta}
    \eeq
    \beq
    y_{\mathrm{M}}(z)=\frac{1}{2\pi i}\oint d\zeta\, \zeta^{-(M+1)}\xi (\zeta ,z)\; ,
    \label{inv_zeta}
    \eeq
and
    \beq
    \xi ''(\zeta ,z)+k^2_y\xi(\zeta ,z)  = aF(\zeta) \xi
    (\zeta ,z)\; .
    \label{xiEQzeta}
    \eeq
When expressing a function of $\theta$, for example the function
$F(\theta)$, in terms of $\zeta$, we write $F(\zeta )= F(\theta )$
above instead of creating a new symbol; this should not introduce
any unnecessary confusion. We adopt this convention throughout
this paper. The singularity of $F(\theta)$ at $\theta =0$,
corresponds to a singularity of $F(\zeta)$ at $\zeta=1$. The
singular part of $F(\zeta)$ is
    \beq
    F(\zeta )\cong \sqrt{\frac{\pi}{1-\zeta}} \hspace{0.5 cm} \mbox{for}
    \hspace{0.2 cm}\zeta\rightarrow 1\; .
    \label{F}
    \eeq

Equation (\ref{eq:xizeta}) is a power series expansion of the
function $\xi$ in the variable $\zeta$. The radius of the
convergence circle of this series is 1, since the closest
singularity of $\xi$ is at $\zeta =1$, i.e., at $\theta=0$. From
the residue theorem, Eq. (\ref{inv_zeta}) is clearly the inverse
of Eq. (\ref{eq:xizeta}) provided that the integration contour
lies inside of the convergence circle, and the contour encircle
the origin $\zeta =0$ counterclockwise. It is convenient to take
the contour to be the unit circle and the singularity to be
located at $\zeta=1+\epsilon$ with a small and positive
$\epsilon$. On the $\zeta$-plane, we make a cut on the real axis
from $\zeta=1+\epsilon$ to $\zeta=\infty$, and make all the
following calculation on the first sheet of the Riemann surface.
Expressed in the $\theta$ variable in Eq. (\ref{xi_inv}), the
singularity is at $\theta = -\,i\,\log(1+\epsilon)$, i.e., below
the contour of Eq. (\ref{xi_inv}). The cut on the $\theta$-plane
is at the lower-half of the imaginary axis, i.e., $\theta$ from
$-i\epsilon$ to $-i\infty$.

We solve in the next section the transient bbu problem by relating
it to the coherent solutions given by Eqs. (\ref{kc}) and
(\ref{eigenSol}).

\section{Initial Value Problem}\label{SecInitial}
One can carry out the bbu calculations in terms of either the
$\zeta$ or the $\theta$ variable. We choose to use the variable
$\zeta$ here. (The paper \cite{WW02} is carried out in the
variable $\theta$.)

We show in this section that the transient solution to the
equation of motion (\ref{eq:motion}) is
    \bea
    y_{\mathrm{M}}(z)
    &=&
    y_{\mathrm{M0}}\cos (k_yz)+y'_{\mathrm{M0}}\sin (k_yz) \;/k_y
    \nonumber\\
    &+&
    \frac{1}{2\pi i}\sum_{N=0}^{M-1}y_{\mathrm{N0}}\oint d\zeta\,\zeta^{-(
    M-N+1)}\cos [k_c(\zeta )z]
    \nonumber\\
    &+&
    \frac{1}{2\pi i}\sum_{N=0}^{M-1}y'_{\mathrm{N0}}\oint d\zeta\,\zeta^{-(
    M-N+1)}\frac{\sin [k_c(\zeta )z]}{k_c(\zeta )}
    \;  ,
    \nonumber \\
    \label{ytrans}
    \eea
where $y_{M0}$ and $y'_{M0}$ are, respectively, the initial values
(values at $z=0$) of $y_{\mathrm{M}}(z)$ and $y_{\mathrm{M}}'(z)$.

First, we find the transient solution of Eq. (\ref{xiequation}).
This equation yields
 \beq\label{xitExplicit}
    \tilde{\xi}(\zeta,s)=\frac{s\,\xi(\zeta,0)+\xi'(\zeta,0)}{s^2+k_y^2-a\,
    F(\zeta)}\; ,
 \eeq
 where
    \beq\label{xitilde}
    \tilde{\xi}(\zeta,s)\equiv \int^{\infty}_0dz\,\xi(\zeta,z)\,e^{-s z}.
    \eeq
 After carrying out the inverse Laplace transform of (\ref{xitilde}), using
 (\ref{xitExplicit}), we obtain
\beq\label{xisolution}
    \xi(\zeta,z)    =
    \xi(\zeta,0)\cos [k_c(\zeta)z]+\xi'(\zeta,0)\frac{\sin [k_c({\zeta})z]}
    {k_c({\zeta})}\;    .
\eeq

In order to obtain (\ref{ytrans}), we substitute the above result
(\ref{xisolution}) into (\ref{inv_zeta}) and then use $\xi
(\zeta,0)=\sum_{M=0}^{\infty}y_{\mathrm{M,0}}\zeta^M$ and
$\xi'(\zeta ,0)=\sum_{M=0}^{\infty}y'_{\mathrm{M,0}}\zeta^M$. The
result is (\ref{ytrans}).  We shall apply the solution
(\ref{ytrans}) to some specific cases in the next section.

\section{Initial Single Bunch Offset}\label{OneOffset}
We study in this section the equation (\ref{ytrans}) for the case
where only the first bunch, i.e. $M=0$, is initially offset
transversely from the center of the chamber,
$y_{\mathrm{M0}}=y_{\mathrm{00}} \delta_{\mathrm{M,0}}$, and
$y'_{\mathrm{M0}}=0, \forall\,M$. In this case, Eq. (\ref{ytrans})
becomes,  for $M\neq 0$,
    \bea
    y_{\mathrm{M}}(z)
    &=&
    \frac{1}{2\pi i}y_{\mathrm{00}}\oint d\zeta \zeta^{-(M+1)}
    \cos [k_c(\zeta )z]
    \nonumber \\
    &\equiv&
    \frac{y_{\mathrm{00}}}{4\pi}\left[\eta^{(+)}_{\mathrm{M}}(z)
    +\eta^{(-)}_{\mathrm{M}}(z)\right]\;,
    \label{yM0}
    \eea
where
    \beq
    \eta^{(\pm)}_{\mathrm{M}}(z)\equiv\frac 1 i \oint d\zeta\exp\left\{
    \Psi^{(\pm)}_{\mathrm{M}}(\zeta)\right\}\;,
    \label{etaplusminus}
    \eeq
with
    \beq\label{NFPsiFirst}
    \Psi^{(\pm)}_{\mathrm{M}}(\zeta)=\pm i k_c(\zeta)z
    -(M+1)\log(\zeta)
    \;.
    \eeq
We wish to find the asymptotic behavior of $y_{\mathrm{M}}$ as
given by Eq. (\ref{yM0}) as $M\rightarrow\infty$; we shall use the
well known saddle point method for this purpose.

The asymptotic behavior of the integral (\ref{yM0}) is determined
by the behavior of $\cos [k_c(\zeta )z]$ near $\zeta =1$, or
$\theta=0$, where the phase difference between adjacent bunches
approaches zero. In other words, the saddle point
$\zeta_{saddle}\rightarrow 1$, or equivalently,
$\theta_{saddle}\rightarrow 0$ in the limit of $M\rightarrow
\infty$. The behavior of $\cos [k_c(\zeta )z]$ near $\zeta =1$ is,
from Eq. (\ref{kc}), controlled by the behavior of $F(\zeta )$ in
the same neighborhood, where $F(\zeta)$ is given in Eq. (\ref{F}).
We shall use the approximation for $F(\zeta )$ in Eq. (\ref{F})
throughout the rest of this paper. Combining the last expression
with Eqs. (\ref{kc}), (\ref{xiEQzeta}) and (\ref{F}), we have
    \beq
    k_c(\zeta )=\sqrt{k_y^2-a\sqrt{\pi /(1-\zeta)}}\;  ,
    \label{kc2}
    \eeq
and
  \beq
    \xi ''(\zeta ,z)+k^2_y\xi(\zeta ,z)  = a\sqrt{\frac{\pi}{1-\zeta}} \xi
    (\zeta ,z)\; .
    \label{xiEQapprox}
    \eeq
The last equation together with the equation(\ref{inv_zeta}) make
up the basis for the remainder of this section.

We shall carry out below the asymptotic analysis of the following
two cases:
\newline
{\bf First Case:} This is the case where either $k_y=0$, or $M$ is
so large that the $a\sqrt{\pi /(1-\zeta)}$ term dominate over
$k_y^2$ in Eq. (\ref{kc2}). As a consequence, we can use the
approximate expression
    \beq
    k_c(\zeta )=a_1i(1-\zeta )^{-1/4}
    \label{kc3}
    \eeq
where $a_1=\sqrt{a\sqrt{\pi}}$. This case will be referred to as
the no focusing case. Clearly, in order for this approximation to
be valid, the condition $|a_1(1-\zeta_{\mathrm{NF}})^{-1/4}|\gg
k_y$, has to be satisfied, where $\zeta_{\mathrm{NF}}$ is the
saddle point.
\newline
{\bf Second Case:} This is the case where $M$ is so large that Eq.
(\ref{F}) is valid, and yet $k_y^2$ in Eq. (\ref{kc2}) dominates
over the $a\sqrt{\pi /(1-\zeta)}$ term. As a consequence,
    \beq\label{k_c_SF}
    k_c(\zeta )\cong k_y-2a_2(1-\zeta)^{-1/2} \;  ,
    \eeq
where $a_2=a\sqrt{\pi}/(4k_y)$. We shall refer to this case as the
strong focusing case. The condition for the validity of this
approximation is  $k_y\gg |a_1(1- \zeta_{\mathrm{SF}})^{-1/4}|$,
where $\zeta_{\mathrm{SF}}$ is the saddle point.

The remainder of this section is devoted to detailed treatment of
these two cases.

\subsection{No Focusing (NF) case}\label{OneOffsetA}

We wish to carry out the saddle point analysis to the integrals
(\ref{yM0}) and (\ref{etaplusminus}) with
    \beq
    \Psi^{(\pm)}_{\mathrm{M}}(\zeta)=(M+1)[\mp 4\alpha_1(1-\zeta)^{-1/4}-
    \log(\zeta)]\;,
    \eeq

    \beq
    \dot{\Psi}^{(\pm)}_{\mathrm{M}}(\zeta)=(M+1)[\pm \alpha_1(1-\zeta)^{-5/4}-
    1/\zeta]\;,
    \eeq
and
    \beq
    \ddot{\Psi}^{(\pm)}_{\mathrm{M}}(\zeta)=(M+1)[\mp(5/4)
    \alpha_1(1-\zeta)^{-9/4}+
    1/\zeta^2]\;,
    \eeq
where ``$\cdot$'' stands for $d/d\zeta$, and ${\alpha}_1\equiv a_1
z/[4(M+1)]$. The function $ \Psi^{(\pm)}_{\mathrm{M}}(\zeta)$ has
branch points at $\zeta=0$ and $\zeta=1$. Let us draw cuts in the
$\zeta$ plane from $\zeta=-\infty$ to $0$, and from $\zeta=1$ to
$\infty$. The integral (\ref{yM0}) is performed on the first sheet
of $ \Psi^{(\pm)}_{\mathrm{M}}(\zeta)$ which is defined to be the
sheet where $ \Psi^{(\pm)}_{\mathrm{M}}(\zeta)=$ real for
$0<\zeta<1$.

The saddle point $\zeta_{\mathrm{NF}}$ satisfies
$\dot{\Psi}^{(\pm)}_{\mathrm{M}}(\zeta_{\mathrm{NF}})=0$, or
    \beq\label{SP_Eq_NF0}
    (1-\zeta_{\mathrm{NF}})^{5/4}=\pm {\alpha}_1\zeta_{\mathrm{NF}}\;.
    \eeq
This equation can not be solved algebraically.  However noting
that ${\alpha}_1=O(1/M)$ is small in the limit of
$M\rightarrow\infty$, we solve the equation by perturbation. In
terms of the variable $\tilde{\zeta}\equiv 1-\zeta$, Eq.
(\ref{SP_Eq_NF0}) becomes, to the lowest order in $\alpha_1$
    \beq\label{SP_Eq_NF0_tilde}
    \tilde{\zeta}_{\mathrm{NF}}^{5/4}=\mp\alpha_1 \;  .
    \eeq
Taking the fourth power of this equation, we have
    \beq\label{SP_Eq_NF0_tilde4}
    \tilde{\zeta}_{\mathrm{NF}}^5=\alpha_1^4 \;  ,
    \eeq
yielding the solutions
    \beq\label{NFSP}
    \tilde{\zeta}_{\mathrm{NF}}={\alpha}_1^{4/5}(1,e^{\pm i2\pi/5},e^{\pm
    i4\pi/5})\;.
    \eeq
The condition (\ref{SP_Eq_NF0_tilde4}) is a necessary but not a
sufficient condition for saddle points, (for example, we took the
fourth power of Eq. (\ref{SP_Eq_NF0_tilde}) in order to obtain Eq.
(\ref{SP_Eq_NF0_tilde4}), we might in doing so have introduced
spurious solutions.) Each of the solutions (\ref{NFSP}) has yet to
be verified to be a relevant saddle point. It is straightforward
to verify that $\zeta_{\mathrm{NF}}^{(-)}=1-{\alpha}_1^{4/5}$ is
the only saddle point of $\eta^{(-)}$, and
$\zeta_{\mathrm{NF}}^{(+)}=1-\alpha_1^{4/5}e^{\pm i4\pi/5}$ are
the only saddle points of $\eta^{(+)}$ we have to consider.

The saddle point contribution to $\eta^{(\pm)}$ satisfies
    \beq
    \eta^{(\pm)}_M\propto\exp\left[\Psi^{(\pm)}
    \left(\zeta_{\mathrm{NF}}^{(\pm)}\right)\right]\;.
    \eeq
Routine calculation gives the following results for the exponents:
   \beq\label{asymEtaMinus}
    \Psi^{(-)}_{\mathrm{M}}\left(\zeta_{\mathrm{NF}}^{(-)}\right)
    = 5(M+1){\alpha}_1^{4/5}
    \;,
    \eeq
and
    \beq\label{asymEtapPlus}
    \Psi^{(+)}_{\mathrm{M}}\left(\zeta_{\mathrm{NF}}^{(+)}\right)
    = 5(M+1){\alpha}_1^{4/5}\exp(\pm i4\pi/5)
    \;.
    \eeq
Notice that the real part of $\Psi^{(+)}$ above is negative;
therefore, $ \eta^{(+)}_M \rightarrow 0$ in the limit of
$M\rightarrow \infty$. We shall ignore the $ \eta^{(+)}_M $ term
in Eq (\ref{yM0}).

In order to perform the saddle point integral for $\eta^{(-)}_M$
we need, in addition to (\ref{asymEtaMinus}), the following
    \beq\label{asymddotEtaMinus}
    \ddot{\Psi}^{(-)}_{\mathrm{M}}\left(\zeta_{\mathrm{NF}}^{(-)}\right)
   =\frac{5(M+1)}{4{\alpha}_1^{4/5}}
    \;.
    \eeq
We notice that
$\ddot{\Psi}^{(-)}_{\mathrm{M}}\left(\zeta_{\mathrm{NF}}^{(-)}\right)
\propto{\alpha}_1^{-9/5}\propto M^{9/5}\rightarrow \infty$ very
fast, as $M\rightarrow\infty$. Such sharp dependence of the
integrand of (\ref{etaplusminus}) in the neighborhood of the
saddle saddle point validates the saddle point approximation.

From the above discussion, the equation
    \beq\label{NFy(-)}
    y_{\mathrm{M}}(z)
    =\frac{y_{\mathrm{00}}}{4\pi}\eta^{(-)}_{\mathrm{M}}(z)
    \propto\exp\left[\Psi^{(-)}\left(\zeta_{\mathrm{NF}}^{(-)}\right) \right]\;
    \eeq
together with Eqs. (\ref{asymEtaMinus}) and
(\ref{asymddotEtaMinus}) are all we need for the saddle point
estimate of the present bbu problem. However, before stating the
results, let us have a discussion on the growth time
$t_{\mathrm{NF}}$ of the mode under discussion.

The $M$-th bunch reaches the linac at time $t=M\tau_{\mathrm{B}}$.
The quantity $\alpha_1$ in the expression (\ref{asymEtaMinus}) can
be written in terms of $M$ and $a$. If we replace $M$ or $M+1$
(recall that $M\gg$1) in the resulting $\Psi^{(-)}$ by
$t/\tau_{\mathrm{B}}$, we obtain
    \beq\label{Psi(-)t}
    \Psi^{(-)}_{\mathrm{M}}\left(\zeta_{\mathrm{NF}}^{(-)}\right)
    =\left(\frac{t}{t_{\mathrm{NF}}}\right)^{1/5}    \;,
    \eeq
where the growth time
    \beq\label{GrowthTimeNF}
    t_{\mathrm{NF}}=
    \frac{\tau_{\mathrm{B}}}{4\pi}\left(\frac{4}{5}\right)^5
    \frac{1}{z^4}\frac{1}{a^2}  \;  ,
    \eeq
and the result of the saddle-point integral is
    \bea
    y_{\mathrm{M}}(z)
    &=&\frac{y_{\mathrm{00}}}{4\pi}\eta^{(-)}_{\mathrm{M}}(z)
                          \nonumber \\
    &=&
    \frac {y_{\mathrm{00}}}{5\sqrt{2\pi}}\left(\frac{t_{\mathrm{NF}}}t
    \right)^{9/10}\frac{\tau_{\mathrm{B}}}{t_{\mathrm{NF}}}
    \exp\left\{\left(\frac{t}{t_{\mathrm{NF}}}\right)^{1/5}\right\} \;  .
    \label{NFFirstSolution}
    \eea

So far we have been dealing with the case of a uniformly filled
linac. If the filling is not uniform, (some buckets not filled,)
the above results do not hold. In Section \ref{SecDamping}, we
shall treat an example of such non-uniform case. In order to
facilitate later comparison, let us write Eq. (\ref{GrowthTimeNF})
for $ t_{\mathrm{NF}}$ in another form. Using Eq. (\ref{a}), Eq.
(\ref{GrowthTimeNF}) becomes
    \beq\label{GTNF-1}
    t_{\mathrm{NF}}=
    \frac{\tau_{\mathrm{B}}}{\pi}\frac{16}{5^5}
    \frac{b^6}{z^4}\frac{1}{\delta^2_{\mathrm{skin}}}
    \frac{I^2_{\mathrm{A}}}{I^2_{\mathrm{B}}}  \;  .
    \eeq
For the case of uniform filling, the
$I_{\mathrm{B}}=e\,N_{\mathrm{B}}/\tau_{\mathrm{B}}$ above equals
the locally averaged current $I_{\mathrm{average}}$. Therefore the
above equation can be expressed as
    \beq\label{GTNF-av}
    t_{\mathrm{NF}}=
    \frac{\tau_{\mathrm{B}}}{\pi}\frac{16}{5^5}
    \frac{b^6}{z^4}\frac{1}{\delta^2_{\mathrm{skin}}}
    \frac{I^2_{\mathrm{A}}}{I^2_{\mathrm{average}}}  \;  .
    \eeq
We shall compare later the above expressions (\ref{GTNF-1}) and
(\ref{GTNF-av}) with the corresponding result for a non-uniformly
filled beam.

\subsection{Strong Focusing (SF) case}
The treatment of this case is similar to the NF case. The exponent
of the integrand of the integral (\ref{etaplusminus}) is, for this
case,
    \bea
    \Psi^{(\pm)}_{\mathrm{M}}(\zeta)
    &=&\pm i k_y z
    \nonumber \\
    &\mp& i 2 a_2 z(1-\zeta)^{-1/2}-(M+1)\log(\zeta)\;.
    \label{SF Psi}
    \eea
This function has branch points at $\zeta=0$ and $\zeta=1$. We cut
the complex $\zeta$ plane from $\zeta=-\infty$ to $0$, and from
$\zeta=1$ to $\infty$. The first two derivatives of
$\Psi^{(\pm)}_{\mathrm{M}}(\zeta)$ are
    \beq
    \dot{\Psi}^{(\pm)}_{\mathrm{M}}(\zeta)=\mp i a_2 z(1-\zeta)^{-3/2}
    -\frac{M+1}\zeta\;,
    \eeq
and
    \beq
    \ddot{\Psi}^{(\pm)}_{\mathrm{M}}(\zeta)=\mp i \frac 32 a_2 z(1
    -\zeta)^{-5/2}+\frac{M+1}{\zeta^2}\;.
    \label{SF ddotPsi}
    \eeq
The saddle point condition $\dot{\Psi}^{(\pm)}_{\mathrm{M}}(
\zeta_{\mathrm{SF}})=0$ leads to
    \beq\label{SP_Eq_SF}
    (1-\zeta_{\mathrm{SF}})^{3/2}=\mp i{\alpha}_2\zeta_{\mathrm{SF}}\;,
    \eeq
with ${\alpha}_2\equiv a_2 z/(M+1)$. Since ${\alpha}_2\rightarrow
0$, as $M\rightarrow\infty$, we could again find the saddle points
by a perturbation method. The result is, to the leading order of
$\alpha_2$,
    \beq\label{SP SF}
    \zeta_{\mathrm{SF}}= \left(1- \alpha_2^{2/3}e^{i\pi/3},1+\alpha_2^{2/3},1-
    \alpha_2^{2/3}e^{-i\pi/3}\right)  \;  ,
    \eeq
where we write the solutions of Eq. (\ref{SP_Eq_SF}) as elements
of a $1\times 3$ row matrix.

The equation (\ref{SP SF}) is a necessary but not a sufficient
condition for the saddle points.  Simple algebraic calculations
shows that the first element of the matrix (\ref{SP SF}) is a
 saddle point of $\eta_M^{(-)}$, and that the third element is
a  saddle point of  $\eta_M^{(+)}$. The second element of (\ref{SP
SF}) which is $>1$ and lies on the branch cut is not accessible to
the integration contour.

We need to evaluate $\Psi^{(\pm)}_{\mathrm{M}}$ and
$\ddot{\Psi}^{(\pm)}_{\mathrm{M}}$ at the appropriate saddle
points. They are
    \begin{equation*}
    \Psi^{(+)}_{\mathrm{M}}(\zeta_{\mathrm{SF},3})
    =
    +i k_y z+ 3(M+1){\alpha}_2^{2/3}\exp\left\{- \frac{i\pi}3
    \right\}\;,
    \end{equation*}
    \begin{equation*}
    \ddot{\Psi}^{(+)}_{\mathrm{M}}(\zeta_{\mathrm{SF},3})
    =\frac{3(M+1)}{2{\alpha}_2^{2/3}}\exp\left\{\frac{i\pi}3
    \right\}
    \;,
    \end{equation*}
    \begin{equation*}
    \Psi^{(-)}_{\mathrm{M}}(\zeta_{\mathrm{SF},1})
    =
    -i k_y z+ 3(M+1){\alpha}_2^{2/3}\exp\left\{\frac{i\pi}3
    \right\}\;,
    \end{equation*}
and
    \begin{equation*}
    \ddot{\Psi}^{(-)}_{\mathrm{M}}(\zeta_{\mathrm{SF},1})
    =\frac{3(M+1)}{2{\alpha}_2^{2/3}}\exp\left\{-\frac{i\pi}3
    \right\}
    \;.
    \end{equation*}
Using these results, we obtain the following asymptotic result for
the displacement of the $M$-th bunch:
    \bea\label{SFSP}
    y_{\mathrm{M}}(z) &=&
    \frac {2y_{\mathrm{00}}}{3\sqrt{2\pi}}\left(
    \frac{t_{\mathrm{SF}}}t\right)^{5/6}\frac{\tau_{\mathrm{B}}}
    {t_{\mathrm{SF}}}\exp\left\{\left(\frac t{t_{\mathrm{SF}}}
    \right)^{1/3}\right\}   \nonumber   \\
    &\times&
    \cos\left[\sqrt{3}\left(\frac t{t_{\mathrm{SF}}}
    \right)^{1/3}-k_yz+\frac{\pi}6\right]\;,
    \eea
where the growth time for this mode
    \beq\label{GrowthTimeSF}
    t_{\mathrm{SF}}\equiv \tau_{\mathrm{B}}\left(
    \frac23 \right)^3\frac 1{a_2^2z^2}\;,
    \eeq
and again $t=(M+1)\tau_{\mathrm{B}}$, or $M\tau_{\mathrm{B}}$
since $M$ is large

The results of this section have been applied to the parameters of
PERL in Ref \cite{WW02}. The conclusion of that study is that the
PERL beam as designed can not survive the resistive wall bbu
without feedback dampers.

\section{Beam with Periodic Gaps}\label{SecDamping}
The bunch filling pattern considered in this section is as
follows: The beam is made of repetitive identical sequences where
each sequence consists of $p$ adjacent filled buckets followed by
$q$ empty buckets; there are in total $r=p+q$ buckets in a
sequence.
\subsection{Equations of motion}
If all the buckets are filled, then
    \beq
    \hat{O}\,y_M\equiv\left(\frac {d^2}{dz^2}+k_y^2\right)y_M
    =\sum^{M-1}_{N=1}S(M-N)\,y_N\;,
    \eeq
where, $S(M)$ is the wake function given in Section \ref{SecEOM};
$S(M)=0$ for $M\leq 0$, and $S(M-N)=a/\sqrt{M-N}$ for $M>0$. The
parameter $a$ is given by Eq. (\ref{a}).
 Note that we have made here a slight change of
convention. We designated the bunches as $M=1,2,3,...$ above
instead of $M=0,1,2,...$ as was done in Section \ref{SecEOM}. We
adopt this new convention throughout this section.

We have to generalize the above equation to include the case of a
beam with periodic empty buckets. Let us use the notation $u
=1,2,3...$ for the sequence number, and $m=1,2....p$ for the bunch
number in a sequence. It is convenient to define, corresponding to
each $u$ a $p\times p$ matrix $\Sm^{(u)}$ with its elements given
by
    \beq
    \Sm^{(u)}_{m,n}=\Sm^{(u)}_{m-n}= S[ur+(m-n)]\;  ,
    \eeq
where  the range of $u$ for $\Sm^{(u)}$ is $u=0,1,2,......$.
Corresponding to the above matrix, we define $1\times p$ column
vector
     \beq
    Y^{(u)}\equiv
    \left(
    \begin{array}{c}
    y_{u,1} \\
    y_{u,2} \\
    \vdots \\
    y_{u,p}
    \end{array}
    \right)\;
    \eeq
where $y_{u,m}$ is the transverse displacement of the $m$-th bunch
in the $u$-th sequence.

The equation of motion for a beam with periodic gaps can now be
written in a compact form similar to Eq. (\ref{eq:motion}),
    \beq\label{EOMmatrix}
    \hat{O}Y^{(u)}=\sum^u_{v=1}\Sm^{(u-v)}Y^{(v)}\;.
    \eeq
We solve this equation in the next subsection.

\subsection{Solutions}
The $m$-th component of the equation of motion (\ref{EOMmatrix})
is
    \beq\label{EOMum}
    \hat{O}y_{u,m}=\sum^u_{v=1}\sum^p_{n=1}\Sm^{(u-v)}_{m-n}y_{v,n}
    \;.
    \eeq
The following generalization of Eqs. (\ref{Ftheta}) and
(\ref{eq:xi}) is convenient:
    \bea\label{Fourierxim}
    \xi_m(\zeta)=
    \sum^{\infty}_{u=1}\zeta^uy_{u,m}\;,    \\
    \Delta_m(\zeta)=\sum^{\infty}_{u=0} \Sm^{(u)}_m \zeta^u
    \;.
    \label{Delta}
    \eea
Then the above three Eqs. (\ref{EOMmatrix}) $\sim$ (\ref{Delta})
lead to
    \beq\label{CompactEOM}
    \hat{O}\xi_m(\zeta)=\sum^p_{n=1}\Delta_{m-n}(\zeta)\xi_n(\zeta)\;.
    \eeq
Once the solution of the last equation is found, the displacement
of the individual bunch is found by substituting the solution into
the inverse of (\ref{Fourierxim}); namely,
    \beq\label{y_u_m}
    y_{u,m}=\frac 1{2\pi i}\oint d\zeta\, \zeta^{-(u+1)}
    \xi_m(\zeta)\;.
    \eeq

The method we use to solve Eq. (\ref{CompactEOM}) is a
generalization of the method of Section \ref{OneOffset}. Note that
for $u \rightarrow\infty$, the contribution to the integral
(\ref{y_u_m}) is dominated by the behavior of the integrand near
$\zeta =1$. Therefore we shall, in analogy to what we did in
Section \ref{OneOffset}, approximate $\Delta_m$ by its singular
part near $\zeta =1$. The singular part is from \cite{zeta}
    \beq
    \Delta_m(\zeta )\cong
    \frac{a}{\sqrt{r}}\sqrt{\frac{\pi}{1-\zeta}} \hspace{0.5 cm}
    \forall \hspace{0.3cm}m \;  ,
    \label{DeltaApprox}
    \eeq
and the corresponding approximation to Eq. (\ref{CompactEOM}) is
    \beq\label{xiEOMgap}
    \hat{O}\,\xi_m(\zeta)\cong\frac{ap}{\sqrt{r}}
    \sqrt\frac{\pi}{1-\zeta}\,\xi_m(\zeta)\;,\hspace{0.5 cm}
    \forall \hspace{0.3cm}m \;  .
    \eeq
This equation together with Eq. (\ref{y_u_m}) gives us the
asymptotic behavior, $u\rightarrow\infty$, of $y_{u,m}$.

Observe the similarity of Eqs. (\ref{xiEOMgap}) and (\ref{y_u_m})
above to the following equations we obtained earlier for the
uniform filling case, (Eqs. (\ref{xiEQapprox}) and
(\ref{inv_zeta}),)
    \bea
    \hat{O}\xi(\zeta,z)=a\sqrt{\frac{\pi}{1-\zeta}} \xi
    (\zeta ,z)\; ,
    \label{uniform-1}   \\
     y_{\mathrm{M}}(z)=\frac{1}{2\pi i}\oint d\zeta\, \zeta^{-(M+1)}\xi (\zeta ,z)\;
     .
    \label{uniform-2}
    \eea
The variable $m$ appears as a passive parameter in Eqs.
(\ref{xiEOMgap}) and (\ref{y_u_m}). Also, these equations can be
obtained from Eqs. (\ref{uniform-1}) and (\ref{uniform-2}) by the
following substitutions:
    \bea
    M,\mbox{ or }(M+1)  \rightarrow&   u \;,
    \label{sub1}    \\
    a   \rightarrow ap/\sqrt{r}    \;  .
    \label{sub2}
    \eea
Therefore, we can obtain the results for Eqs. (\ref{xiEOMgap}) and
(\ref{y_u_m}) from the corresponding results for the uniform
filling case. We treat here the ``No Focusing '' case
corresponding to the subsection \ref{OneOffset}-A. We specifically
consider the growth time $t^{gap}_{\mathrm{NF}}$ for the beam with
periodic gaps. The ``Strong Focusing '' case can be treated in a
similar way.

We start from the exponent $\Psi^{(-)}_{\mathrm{M}}$ as given by
(\ref{asymEtaMinus}). Expressing $\alpha_1$ in terms of $a$, this
equation is equivalent to
    \beq\label{Psi-a}
    \Psi^{(-)}_{\mathrm{M}}\left(\zeta_{\mathrm{NF}}^{(-)}\right)
    = 5\pi^{1/5}(M+1)^{1/5}(z/4)^{4/5}a^{2/5}
    \;.
    \eeq
Now applying the substitution rules (\ref{sub1}) and (\ref{sub2})
to Eq. (\ref{Psi-a}), we obtain
    \beq\label{PsiGap}
    \Psi^{(-)}_{\mathrm{gap},u}\left(\zeta_{\mathrm{NF}}^{(-)}\right)
    = \left(\frac{u}{u_{\mathrm{NF}}}\right)^{1/5}  \;  ,
    \eeq
where
    \beq\label{uNF}
    u_{\mathrm{NF}}=\frac{1}{4\pi}\left(\frac{4}{5}\right)^5
    \frac{1}{z^4}\frac{r}{a^2p^2}
    \eeq
is the growth time in units of sequences.

We have to translate $u$ into time $t$. The bunch $(u,m)$ reaches
the linac at $t=(ur+m)\tau_{\mathrm{B}}\cong ur\tau_{\mathrm{B}}$.
Therefore we should set $u\rightarrow t/r\tau_{\mathrm{B}}$ and
    \bea
    t^{\mathrm{gap}}_{\mathrm{NF}}&=&r\tau_{\mathrm{B}}u_{\mathrm{NF}}\\
    &=&\frac{\tau_{\mathrm{B}}}{4\pi}\left(\frac{4}{5}\right)^5
    \frac{1}{z^4}\frac{1}{a^2}\frac{r^2}{p^2}   \\
    &=&\frac{\tau_{\mathrm{B}}}{\pi}\frac{16}{5^5}
    \frac{b^6}{z^4}\frac{1}{\delta^2_{\mathrm{skin}}}
    \frac{I^2_{\mathrm{A}}}{I^2_{\mathrm{B}}} \frac{r^2}{p^2}\; .
    \eea
These expressions differ from Eq. (\ref{GrowthTimeNF}) or
(\ref{GTNF-1}) by a factor of $r^2/p^2$. However, this difference
is superficial. Let us calculate the average current of a
sequence. It is clearly
    \beq
    I_{\mathrm{average}}=\frac{p}{r}I_{\mathrm{B}}  \;  .
    \eeq
In terms of $I_{\mathrm{average}}$, the growth time becomes
    \beq
    t^{\mathrm{gap}}_{\mathrm{NF}}=\frac{\tau_{\mathrm{B}}}{\pi}\frac{16}{5^5}
    \frac{b^6}{z^4}\frac{1}{\delta^2_{\mathrm{skin}}}
    \frac{I^2_{\mathrm{A}}}{I^2_{\mathrm{average}}} \; .
    \eeq
This is identical to (\ref{GTNF-av}). We therefore conclude that
the coupled-bunch resistive-wall bbu is a locally-averaged current
effect.

\section{Injection Error and Screen Effect}\label{SecAll}
In this section, we study Eq. (\ref{ytrans}) for the case where
bbu is started up by an injection error, i.e., all the bunches are
initially offset by the same amount,
$y_{\mathrm{M0}}=y_{\mathrm{00}}$, and $y'_{\mathrm{M0}}=0,
\forall\,M\geq 0$. Then, Eq. (\ref{ytrans}) becomes
    \bea
    y_{\mathrm{M}}(z)
    &=&
    y_{\mathrm{00}}\cos(k_yz)
    \nonumber \\
    &-&
    \frac{1}{2\pi i}y_{\mathrm{00}}\oint d\zeta \zeta^{-1}
    (1-\zeta)^{-1}\cos k_c(\zeta)z
    \nonumber \\
    &+&
    \frac{1}{2\pi i}y_{\mathrm{00}}\oint d\zeta \zeta^{-(M+1)}
    (1-\zeta)^{-1}\cos k_c(\zeta)z
    \label{yM0All-1}
    \nonumber \\
    &=&
    y_{\mathrm{00}}\cos(k_yz)-y_{\mathrm{00}}\cos(k_c(0)z)
    \nonumber \\
    &+&
    \frac{y_{\mathrm{00}}}{4\pi}\left[\eta^{(+)}_{\mathrm{M}}(z)
    +\eta^{(-)}_{\mathrm{M}}(z)\right]  \;  ,
    \label{yM0All-2}
    \eea
where, $\eta^{(\pm)}_{\mathrm{M}}(z)$ is given by Eq.
(\ref{etaplusminus}) with
    \beq\label{NFPsiAll}
    \Psi^{(\pm)}_{\mathrm{M}}(\zeta)=\pm i k_c(\zeta)z -(M+1)\log
    (\zeta)
    -\log(1-\zeta)\;.
    \eeq
 Compared with Eq. (\ref{NFPsiFirst}), Eq.
(\ref{NFPsiAll}) has an additional term $-\log(1-\zeta)$ on the
right hand side. We shall see presently that this term does not
change the eigen solutions as given in Secs \ref{SecEOM} $\sim$
\ref{SecDamping}, but it will change the transient solutions. We
shall also see that this term leads to an interesting ``Screen
Effect''. From $k_c(0)=k_y$, Eq. (\ref{yM0All-2}) becomes
simplified to
    \beq
    y_{\mathrm{M}}(z)=\frac{y_{\mathrm{00}}}{4\pi}\left[\eta^{(+)}_{\mathrm{M}}(z)
    +\eta^{(-)}_{\mathrm{M}}(z)\right]\;    .
    \label{yM0All-siplified}
    \eeq
Let us discuss as before two extreme cases: the No Focusing case
and the Strong Focusing case.

\subsection{No Focusing (NF) case}
Similar to what was done in Sec. \ref{OneOffset}, we wish to carry
out the saddle point analysis to the integral (\ref{etaplusminus})
with the exponent
    \beq
    \Psi^{(\pm)}_{\mathrm{M}}(\zeta)=\mp a_1 z(1-\zeta)^{-1/4}-(M+1)
    \log(\zeta)-\log(1-\zeta)\; .
    \label{exponent-all}
    \eeq
The first two derivatives of the exponent are
    \beq
    \dot{\Psi}^{(\pm)}_{\mathrm{M}}(\zeta)=\mp \frac 14 a_1 z(1-\zeta)^{-5
    /4}-\frac{M+1}\zeta+\frac1{1-\zeta}\;,
    \eeq
and
    \beq\label{NFAllPsiDDot}
    \ddot{\Psi}^{(\pm)}_{\mathrm{M}}(\zeta)=\mp \frac 5{16} a_1 z(1
    -\zeta)^{-9/4}+\frac{M+1}{\zeta^2}+\frac 1{(1-\zeta)^2}\;.
    \eeq

The saddle points are determined by $\dot {\Psi}^{(\pm)}
_{\mathrm{M}} (\zeta_{\mathrm{NF}})=0$, i.e.,
    \beq\label{NFALLSP}
    0=\mp \frac 14 a_1 z(1-\zeta_{\mathrm{NF}})^{-5
    /4}-\frac{M+1}{\zeta_{\mathrm{NF}}}+\frac1{1-\zeta_{\mathrm{NF}}}\;,
    \eeq
which can not be solved algebraically. However, since the saddle
points $\zeta_{\mathrm{saddle}}\rightarrow 1$ in the limit of
$M\rightarrow\infty$, we could solve Eq. (\ref{NFALLSP}) by a
perturbation method. In terms of $\tilde{\zeta}\equiv 1- \zeta$,
Eq. (\ref{NFALLSP}) becomes
    \beq\label{NFFull}
    \mp\frac 14 a_1 z \tilde{\zeta}_{\mathrm{NF}}^{-5/4}
    \pm\frac 14
    a_1 z \tilde{\zeta}_{\mathrm{NF}}^{-1/4}+
    \tilde{\zeta}_{\mathrm{NF}}^{-1}-1=M+1\;.
    \eeq
Keeping the leading term in Eq. (\ref{NFFull}), we get
    \beq\label{NFFullFirstOrder}
    \mp\frac 14 a_1 z \tilde{\zeta}_{\mathrm{NF}}^{-5/4}=M+1\;.
    \eeq
The last equation is identical to Eq. (\ref{SP_Eq_NF0_tilde}), and
therefore this yields the same first-order solution given in Eq.
(\ref{NFSP}). We select now the relevant saddle points by
repeating what we did before following Eq. (\ref{NFSP}), and then
carry out the saddle point integral corresponding to the exponent
(\ref{exponent-all}). The result is
    \bea\label{NFAllSolution}
    y_{\mathrm{M}}(z)
    &=&
    \mathcal{G}_{\mathrm{NF}}
    \frac {y_{\mathrm{00}}}{5\sqrt{2\pi}}\left(\frac{t_{\mathrm{NF}}}t
    \right)^{9/10}\frac{\tau_{\mathrm{B}}}{t_{\mathrm{NF}}}\exp\left\{
    \left(\frac t{t_{\mathrm{NF}}}\right)^{1/5}\right\}\;,
    \nonumber \\
    \eea
where the growth time
    \beq\label{GT-NF-all}
    t_{\mathrm{NF}}=
    \frac{\tau_{\mathrm{B}}}{4\pi}\left(\frac{4}{5}\right)^5
    \frac{1}{z^4}\frac{1}{a^2}  \;  ,
    \eeq
and
    \bea
    \mathcal{G}_{\mathrm{NF}}
    &\equiv&
    5\left(\frac t{\tau_B}\right)^{4/5}
    \left(\frac {t_{\mathrm{NF}}}{\tau_B}\right)^{1/5}
    \nonumber \\
    &=&
    4\left(\frac1{4\pi a^2}\right)^{1/5}\left(\frac Mz\right)^{4/5}\;.
    \eea
It is very interesting to compare with the above result
(\ref{NFAllSolution}) to the result (\ref{NFFirstSolution}) of the
initial single-bunch offset case. (1) The growth time
$t_{\mathrm{NF}}$ is the same for both cases, as it should be,
since $t_{\mathrm{NF}}$ should depends only on the eigen
solutions. (2) The only difference between the transient solutions
is the factor $\mathcal{G}_{\mathrm{NF}}$ which is proportional to
$M^{4/5}$ instead of to $M$, (recall that $t\propto M$.) This is
surprising: Since $\theta_{saddle}\cong 0$, we would expect all
the bunches preceding the bunch $M$ to excite this bunch by the
same amount leading to $\mathcal{G}_{\mathrm{NF}}\propto M$.
Clearly, the preceding bunches are screening each other. (It can
actually be shown that for large but not too large $M$, the
function $\mathcal{G}_{\mathrm{NF}}\propto M$.)

\subsection{Strong Focusing (SF) case}
Let us not go into detailed discussion of this case, since the
arguments are so similar to those of Section \ref{OneOffset} and
Section \ref{SecAll}-A. We just list the results:

    \bea\label{SFAllSolution}
    y_{\mathrm{M}}(z)
    &=&
    \mathcal{G}_{\mathrm{SF}}
    \frac {2y_{\mathrm{00}}}{3\sqrt{2\pi}}\left(
    \frac{t_{\mathrm{SF}}}t\right)^{5/6}\frac{\tau_{\mathrm{B}}}
    {t_{\mathrm{SF}}}\exp\left\{\left(\frac t{t_{\mathrm{SF}}}
    \right)^{1/3}\right\}
    \nonumber \\
    &\times&
    \cos\left[\sqrt{3}\left(\frac t{t_{\mathrm{SF}}}
    \right)^{1/3}-k_yz-\frac{\pi}6\right]\;,
    \eea
where
    \beq\label{GTSF-all}
    t_{\mathrm{SF}}\equiv \tau_{\mathrm{B}}\left(
    \frac23 \right)^3\frac 1{a_2^2z^2}\;,
    \eeq
and
    \bea
    \mathcal{G}_{\mathrm{SF}}
    &\equiv&
    \frac 32\left(\frac t{\tau_B}\right)^{2/3}
    \left(\frac {t_{\mathrm{SF}}}{\tau_B}\right)^{1/3}
    \nonumber \\
    &=&
    \left(\frac{16k_y^2}{\pi a^2}\right)^{1/3} \left( \frac Mz
    \right)^{2/3}\;.
    \eea

\begin{acknowledgments}
The authors thank Professor T.O. Raubenheimer of the Stanford
Linear Accelerator Center for an illuminating comment. This work
was supported by US Department of Energy under contract
DE-AC02-98CH10886 (Jiunn-Ming Wang) and contract DE-AC03-76SF00515
(Juhao Wu).
\end{acknowledgments}

\end{document}